\documentclass[aps,twocolumn,superscriptaddress,showpacs]{revtex4-1}

\usepackage{graphicx}
\usepackage{dcolumn}
\usepackage{bm}
\usepackage{amsmath}
\usepackage{color}
\usepackage{ulem}
\usepackage{threeparttable}
\usepackage{enumerate}
\usepackage{paralist}

\begin{document}

\title{Low-temperature specific heat and heat transport of Tb$_2$Ti$_{2-x}$Zr$_x$O$_7$ single crystals}

\author{H. L. Che}
\affiliation{Department of Physics and Key Laboratory of Strongly-Coupled Quantum Matter Physics (CAS), University of Science and Technology of China, Hefei, Anhui 230026, People's Republic of China}
\affiliation{Department of Physics, Lyuliang University, Lyuliang, Shanxi 033001, People's Republic of China}

\author{S. J. Li}
\affiliation{Hefei National Research Center for Physical Sciences at Microscale, University of Science and Technology of China, Hefei, Anhui 230026, People's Republic of China}

\author{J. C. Wu}
\affiliation{Hefei National Research Center for Physical Sciences at Microscale, University of Science and Technology of China, Hefei, Anhui 230026, People's Republic of China}

\author{N. Li}
\affiliation{Hefei National Research Center for Physical Sciences at Microscale, University of Science and Technology of China, Hefei, Anhui 230026, People's Republic of China}

\author{S. K. Guang}
\affiliation{Department of Physics and Key Laboratory of Strongly-Coupled Quantum Matter Physics (CAS), University of Science and Technology of China, Hefei, Anhui 230026, People's Republic of China}

\author{K. Xia}
\affiliation{Department of Physics and Key Laboratory of Strongly-Coupled Quantum Matter Physics (CAS), University of Science and Technology of China, Hefei, Anhui 230026, People's Republic of China}

\author{X. Y. Yue}
\affiliation{Institute of Physical Science and Information Technology, Anhui University, Hefei, Anhui 230601, People's Republic of China}

\author{Y. Y. Wang}
\affiliation{Institute of Physical Science and Information Technology, Anhui University, Hefei, Anhui 230601, People's Republic of China}

\author{X. Zhao}
\affiliation{School of Physical Sciences, University of Science and Technology of China, Hefei, Anhui 230026, People's Republic of China}

\author{Q. J. Li}
\email{liqj@ahu.edu.cn}
\affiliation{School of Physics and Material Sciences, Anhui University, Hefei, Anhui 230601, People's Republic of China}
\affiliation{Hefei National Research Center for Physical Sciences at Microscale, University of Science and Technology of China, Hefei, Anhui 230026, People's Republic of China}

\author{X. F. Sun}
\email{xfsun@ustc.edu.cn}
\affiliation{Department of Physics and Key Laboratory of Strongly-Coupled Quantum Matter Physics (CAS), University of Science and Technology of China, Hefei, Anhui 230026, People's Republic of China}
\affiliation{Institute of Physical Science and Information Technology, Anhui University, Hefei, Anhui 230601, People's Republic of China}

\date{\today}

\begin{abstract}

We report a study on the specific heat and heat transport of Tb$_2$Ti$_{2-x}$Zr$_x$O$_7$ ($x =$ 0, 0.02, 0.1, 0.2, and 0.4) single crystals at low temperatures and in high magnetic fields. The magnetic specific heat can be described by the Schottky contribution from the crystal-electric-field (CEF) levels of Tb$^{3+}$, with introducing Gaussian distributions of the energy split of the ground-state doublet and the gap between the ground state and first excited level. These crystals has an extremely low phonon thermal conductivity in a broad temperature range that can be attributed to the scattering by the magnetic excitations, which are mainly associated with the CEF levels. There is strong magnetic field dependence of thermal conductivity, which is more likely related to the field-induced changes of phonon scattering by the CEF levels than magnetic transitions or spin excitations. For magnetic field along the [111] direction, there is large thermal Hall conductivity at low temperatures which displays a broad peak around 8 T. At high fields up to 14 T, the thermal Hall conductivity decreases to zero, which supports its origin from either the spinon transport or the phonon skew scattering by CEF levels. The thermal Hall effect is rather robust with Zr doping up to 0.2 but is strongly weakened in higher Zr-doped sample.

\end{abstract}

\maketitle

\section{INTRODUCTION}

The pyrochlore rare-earth titanites have been a focus for the study of the physics of spin frustration and quantum magnets \cite{GardnerRev, GingrasRev}. The phase diagram of pyrochlore magnets with nearest-neighbor exchange and long-range dipolar interactions was given by Hertog {\it et al.} \cite{den} Ho$_2$Ti$_2$O$_7$ and Dy$_2$Ti$_2$O$_7$ with effective ferromagnetic exchange and Ising anisotropy have a disordered spin-ice ground state \cite{Ramirez, Bramwell, PhysRevLett792554}. Tb$_2$Ti$_2$O$_7$ has similar Ising anisotropy to that of Ho$_2$Ti$_2$O$_7$ and Dy$_2$Ti$_2$O$_7$ but exhibits quite different magnetism. The Tb$^{3+}$ ions form a pyrochlore lattice and the spin interactions are antiferromagnetic (AF) ($\theta\rm_{CW}$ = -19 K) \cite{Gingras, Gardner1}. Tb$^{3+}$ is a non-Kramers ion with the angular momentum $J =$ 6, which can have in total 13 crystal-electric field (CEF) levels. The lowest level is a ground-state doublet and the first-excited state doublet is 1.5 meV above \cite{Zhang, Mirebeau}. Note that the estimations of the nearest-neighbor exchange and long-range dipolar interactions yield $J_{nn}$/D$_{nn} \sim -$ 1, placing Tb$_2$Ti$_2$O$_7$ very close to the boundary between the long-rang N\'eel ordered {\bf Q} = 0 phase and the dipolar spin-ice state \cite{den, Gingras}. It was found that Tb$_2$Ti$_2$O$_7$ has no conventional long-range magnetic order down to $\sim$ 0.05 K \cite{Gardner1, Zhang, Mirebeau, JPhysCondensMatter24, PhysRevB68180401}. It does not have the spin-ice ground state at low temperatures also. Instead, some experiments suggested Tb$_2$Ti$_2$O$_7$ having a ground state of quantum spin ice, which is a special type of quantum spin liquids (QSL) \cite{Yin, Molavian, JStatPhys116755, GingrasRev}. However, the elementary excitation of QSL, spinon, has not been confirmed by neutron scattering. At low temperature, pinch points have been observed in the neutron scattering, suggesting that it has a magnetic Coulomb phase governed by the ice rule \cite{Molavian, Fennell1}. In addition, some neutron scattering experiment revealed a novel magnetoelastic mode from the coupling between the transverse acoustic phonons and the excited CEF level, forming a hybrid excitation, which may prevent the magnetic order and the structural distortion \cite{Fennell}. Therefore, the ground state of Tb$_2$Ti$_2$O$_7$, the magnetic excitations and the influence of CEF effects need to be further explored.

Low-temperature heat transport properties can effectively reveal the magnetic excitations of the quantum magnets. For example, the spinons of the QSL candidates can be detected by the ultralow-temperature thermal conductivity measurements, which has the advantage of detecting only the itinerant quasiparticles \cite{Berman, Science328, PhysRevResearch2013099, PhysRevB97104413, NatCommun11, NatCommun12}. Recently, thermal Hall effect has been found to be able to contributed by the magnetic excitations also \cite{PhysRevLett104, Ong, Science329, PhysRevLett115, PhysRevLett121, ProcNatlAcadSci113, Nature559, Science373}. In this regard, it is sometimes difficult to distinguish the contributions from the magnetic excitations and phonons. An interesting finding is that Tb$_2$Ti$_2$O$_7$ exhibits a large thermal Hall effect at low temperatures, which was discussed to be caused by the spinon excitations in the QSL state \cite{Ong}. However, we have found that Tb$_2$Ti$_2$O$_7$ has very small thermal conductivity at low temperatures and attributed it to a phonon-glass-like behavior \cite{Li}, which suggested negligibly weak heat transport of magnetic excitations. Therefore, it is necessary to further investigate the origin the low-temperature thermal Hall effect of this material.

The spin liquid state in Tb$_2$Ti$_2$O$_7$ would be destroyed by the relatively modest perturbations, such as applying magnetic field or doping nonmagnetic impurities. In these cases, the crystal lattice will be changed and the spin fluctuations will be suppressed, which will affect the ground state properties. In this work, we grew the single crystals of Tb$_2$Ti$_{2-x}$Zr$_x$O$_7$ ($x =$ 0, 0.02, 0.1, 0.2, and 0.4) and studied their specific heat, thermal conductivity and thermal Hall effect at low temperatures and in high magnetic fields. We quantitatively analyzed the magnetic specific heat by using the modified Schottky formula, considering a Gaussian distribution of the energy split of the ground-state doublet and the gap between the ground state and first excited level. These crystals show the extremely low phonon thermal conductivity at low temperatures, indicating strong scattering between phonons and magnetic excitations. It was confirmed by the strong magnetic field dependence of $\kappa(B)$, mainly due to the change of phonon scattering by magnetic excitations. The thermal Hall conductivity $\kappa_{xy}(B)$ are large at low temperatures and display a broad peak at 8 T. The disappearance of $\kappa_{xy}(B)$ signal at higher fields demonstrate that it originates from either spinon transport or phonon skew scattering by CEF levels.

\section{EXPERIMENTS}

High-quality Tb$_2$Ti$_{2-x}$Zr$_x$O$_7$ ($x =$ 0, 0.02, 0.1, 0.2, and 0.4) single crystals were grown by using a floating-zone technique \cite{JCrystGrowth191, JCrystGrowth377}. These crystals could be grown well under different oxygen pressures. Tb$_2$Ti$_2$O$_7$ crystal was grown in 0.4 MPa pure oxygen with a growth rate of 2.5 mm/h. In the Zr-doped samples, the incorporation of Zr$^{4+}$ ions with large ionic radius causes the ionic radius of $B$-site (Zr and Ti) to increase. As a result, Zr-doped crystals require smaller oxygen pressure for growing. Meanwhile, with the increase of the doping ratio, the growth rate should be appropriately reduced. The color of these crystals is brown \cite{JCrystGrowth377}. It should be pointed out that the color of Tb$_2$Ti$_2$O$_7$ single crystals grown by different groups exhibit some difference in color, which may be due to the small amount of imperfection or non-stoichiometry \cite{JCrystGrowth191, JCrystGrowth377, Zhang2, Nii, Fennell1, Prabhakaran}

Using X-ray Laue photographs, large pieces of crystals were cut into rectangular shaped samples with specific orientations. The thermal conductivity and thermal Hall conductivity were measured by means of the steady state method at low temperatures down to 0.3 K and in magnetic fields up to 14 T. Heat current was generated by chip resistor and the temperature gradient was probed by by RuO$_2$ thermometers. The thermal conductivity was measured with ``one heater, two thermometers" \cite{NatCommun11, NatCommun12}, while the thermal Hall conductivity was measured with ``one heater, three thermometers" \cite{Ong}. The specific heat was measured by the relaxation method in the temperature range from 0.4 to 30 K using a commercial physical property measurement system (PPMS, Quantum Design) equipped with a $^{3}$He insert. DC magnetic susceptibility ($\chi$) and magnetization were measured using a Quantum Design SQUID-VSM.

\section{RESULTS AND DISCUSSION}

\begin{figure}
\centering\includegraphics[clip,width=6.5cm]{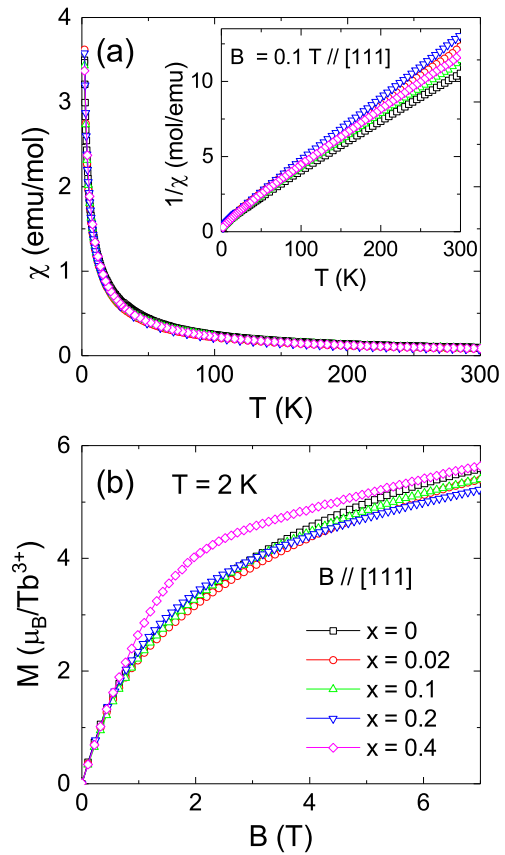}
\caption{(a) Temperature dependence of the magnetic susceptibility of Tb$_2$Ti$_{2-x}$Zr$_x$O$_7$ ($x =$ 0, 0.02, 0.1, 0.2, and 0.4) single crystals in 0.1 T field along the [111] direction. Inset: the temperature dependence of the inverse susceptibility. (b) Magnetization curves of Tb$_2$Ti$_{2-x}$Zr$_x$O$_7$ single crystals at 2 K and with field along the [111] direction.}
\end{figure}

Figure 1(a) shows the temperature dependence of the magnetic susceptibility of Tb$_2$Ti$_{2-x}$Zr$_x$O$_7$ ($x =$ 0, 0.02, 0.1, 0.2, and 0.4) single crystals in 0.1 T field along the [111] direction. The $\chi(T)$ mainly follows the Curie-Weiss behavior except for low temperature region. The magnetization curves at 2 K of these samples are shown in Fig. 1(b), which shows a gradual spin polarization behavior. These results are well known for Tb$_2$Ti$_{2}$O$_7$ \cite{Gingras, Gardner1}. The Zr doping does not change the magnetic susceptibility so much since Zr$^{4+}$ is nonmagnetic.

\begin{figure}
\centering\includegraphics[clip,width=6.5cm]{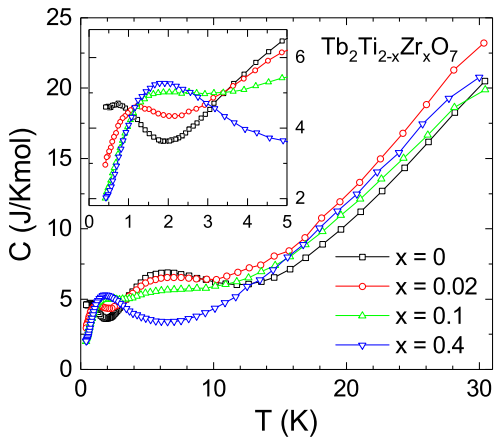}
\caption{Temperature dependence of the low-temperature specific heat of Tb$_2$Ti$_{2-x}$Zr$_x$O$_7$ ($x =$ 0, 0.02, 0.1, and 0.4) single crystals in zero field. Inset: the specific-heat data at low temperature range.}
\end{figure}

Figure 2 shows the zero-field specific heat of Tb$_2$Ti$_{2-x}$Zr$_x$O$_7$ ($x =$ 0, 0.02, 0.1, and 0.4) single crystals at low temperatures. The data of Tb$_2$Ti$_2$O$_7$ has been reported in our earlier work \cite{Li} and display two broad peaks at about 0.7 and 6 K, which could be related to the CEF excitations of Tb$^{3+}$ ions. The ground state and the first excitation of CEF levels are non-Kramers doublets with $\sim$ 18 K separation, and higher levels are singlets. The peak at 6 K is attributed to the low-lying CEF excitations with the energy gap between the ground-state doublet and the first-excited level, and the peak at 0.7 K is attributed to the ground-state doublet, broadened by the exchange interactions. With increasing the Zr doping, the high-$T$ peak is gradually suppressed and disappears, while the low-$T$ peak increases and shifts to higher temperature and becomes stronger. This could be due to the doping effect on the Tb$^{3+}$ CEF levels.

\begin{figure}
\centering\includegraphics[clip,width=6.5cm]{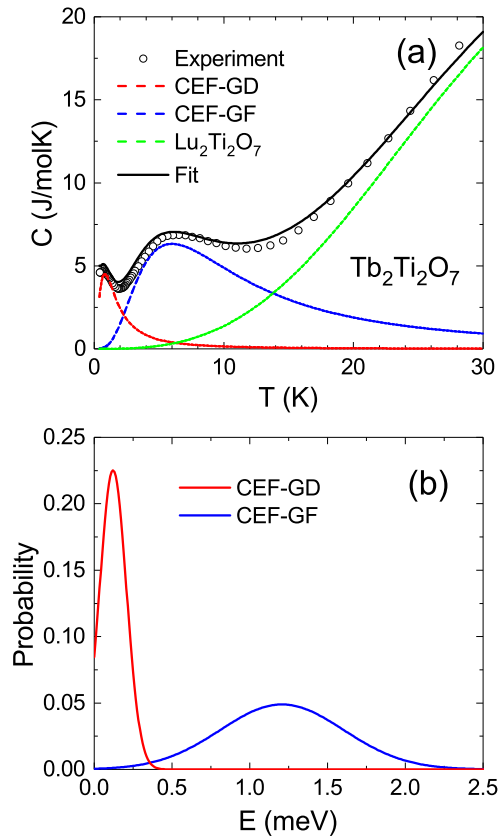}
\caption{(a) Specific heat of Tb$_2$Ti$_2$O$_7$ single crystal. The fit to the experimental data was performed by using the CEF-GD, CEF-GF, and lattice contributions (see the main text), as shown by the red, blue, and green dashed lines, respectively. The sum of these contributions is indicated by the black solid line, which is in good agreement with the experimental results. (b) Distribution probability of the energy gap of the ground-state doublet splitting and the energy gap between the ground state and first-excited state used in the fitting procedure.}
\end{figure}

We analyzed the specific heat data quantitatively. The zero-field specific heat of Tb$_2$Ti$_2$O$_7$ can be fitted by the formula
\begin{equation}\label{eq:eps}
C(T)=C_{L}(T)+2N_{A}I(I+1){\Delta_n}^2 \frac{k_B}{3T^2}+C_{m}(T).
\end{equation}
Here, the first term is the contribution of lattice and the second term is the nuclear Schottky contribution. $N_{A}$ is the Avogadro constant, $I$ is the nuclear spin, $k_B$ is the Boltzmann constant and $\Delta_n$ is the nuclear spin energy splitting, which is independent of magnetic field \cite{Leyarovski, Heltemes, Sanchez, Takatsu}. The second formula is a good approximation for the nuclear Schottky term when the temperatures are not very low, although other formulas can also be used in some cases \cite{Gordon, Hamaguchi}. The last term is the magnetic specific heat. Considering the magnetic specific heat from the CEF levels, we firstly tried to fit it using the standard two-level Schottky formula
\begin{equation}\label{eq:eps}
\begin{split}
C_{m}(T) = a_0R{(\frac{\Delta_0 }{{T}})^2}\frac{{{e^{\Delta_0 /T}}}}{{{{(1 + {e^{\Delta_0 /T}})}^2}}}+a_1R{(\frac{\Delta_1}{{T}})^2}\frac{{{e^{\Delta_1 /T}}}}{{{{(1 + {e^{\Delta_1 /T}})}^2}}},
\end{split}
\end{equation}
where $R$ is the universal gas constant, $\Delta_0$ represents the energy splitting of the Tb$^{3+}$ ground-state doublet, $\Delta_1$ represents the energy gap between the ground state and first excited level, and $a_0$ and $a_1$ are the coefficients, which are equal to 2 because there are two Tb$^{3+}$ ions in the chemical formula. The first and second terms can be defined as CEF-GD and CEF-GF, respectively. However, it is easily found that Eq. (1) and (2) cannot describe the experiment data well. We modified the Schottky formula by considering the distribution of energy splitting, that is, the inhomogeneous distribution of the energy gaps should be taken into account. We used a Gaussian distribution to describe $\Delta_0$ and $\Delta_1$. The specific heat contributed by the Tb$^{3+}$ CEF levels is then given by
\begin{multline}\label{eq:eps}
C_{m}(T) = \int_{0}^{\infty}{\frac{a_0}{{\sqrt {2{\rm{\pi}}{\sigma_{0} ^2}} }}\exp ( {\frac{{ - {{\left( {\Delta - {\Delta _0}} \right)}^2}}}{{2{\sigma_{0} ^2}}}} )}C_{Sch}(T,\Delta)d\Delta
\\+\int_{0}^{\infty}{\frac{a_1}{{\sqrt {2{\rm{\pi}}{\sigma_{1} ^2}} }}\exp ( {\frac{{ - {{\left( {\Delta - {\Delta _1}} \right)}^2}}}{{2{\sigma_{1} ^2}}}} )}C_{Sch}(T,\Delta)d\Delta,
\end{multline}
where $\sigma_{0}$ and $\sigma_{1}$ are the variances, $C_{Sch}(T, \Delta) = 2R{(\Delta / T)^2} e^{\Delta /T}/(1 + e^{\Delta /T})^2$. As shown by the black solid lines in Fig. 3(a), using Eq. (1) and (3) we can fit the specific heat data of Tb$_2$Ti$_2$O$_7$ rather well, where the lattice contribution is estimated from that of nonmagnetic Lu$_2$Ti$_2$O$_7$ \cite{JCrystGrowth377}. The CEF-GD contribution with $\Delta _0$ = 1.5 K and $\sigma_{0}$ = 1.2 K and the CEF-GF contribution with $\Delta _1$ = 14.0 K $\sigma_{1}$ = 4.6 K are shown as the red and blue dashed lines in Fig. 3(a), respectively. The corresponding energy gap distributions are presented in Fig. 3(b). In some previous studies, similarly inhomogeneous doublet splitting has been considered for some non-Kramers pyrochlore materials Pr$_2$Ru$_2$O$_7$, Pr$_{2-x}$Bi$_x$Ru$_2$O$_7$, and Pr$_2$Zr$_2$O$_7$ \cite{Duijn, Tachibana, Wen}.

\begin{figure}
\centering\includegraphics[clip,width=8.5cm]{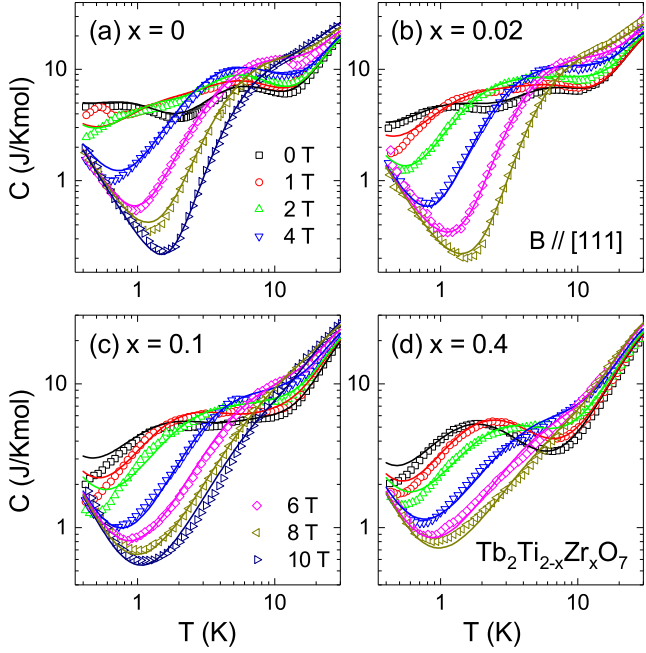}
\caption{ Low-temperature specific heat of Tb$_2$Ti$_{2-x}$Zr$_x$O$_7$ ($x =$ 0, 0.02, 0.1, and 0.4) single crystals in different magnetic fields along the [111] direction. The lines are the fitting results including different contributions.}
\end{figure}

Figure 4 shows the low-temperature specific heat of Tb$_2$Ti$_{2-x}$Zr$_x$O$_7$ ($x =$ 0, 0.02, 0.1, and 0.4) single crystals in different magnetic fields along the [111] direction. The data of Tb$_2$Ti$_2$O$_7$ were taken from our earlier work \cite{Li}. At high magnetic fields, all samples exhibit low temperature ($<$ 1 K) upturn, which is likely the contribution of nuclear spin specific heat. For samples of $x =$ 0 and 0.02, the low-$T$ peak shifts to high temperature and merges with the high-$T$ peak and further shifts to higher temperature with increasing magnetic field. In highly doped sample of $x =$ 0.4, there is only one broad peak that shifts to high temperature and finally disappears with increasing magnetic field.

\begin{table}[htbp]
\caption {The fitting parameters of Eq. (1) and (3), the median of energy gap $\Delta_0$ and $\Delta_1$, the variance $\sigma_0$ and $\sigma_1$ by considering the Gaussian distribution of energy gaps, the nuclear spin energy splitting $\Delta_n$ = 0.12 K, for Tb$_2$Ti$_2$O$_7$ specific heat with increasing magnetic field ($B \parallel$ [111]).} \centering
\renewcommand\arraystretch{1.5}
\begin{tabular}{p{46pt}p{24pt}p{24pt}p{24pt}p{24pt}p{24pt}p{24pt}p{24pt}}
\hline
\hline
  $\mu_0H$ (T) & 0 & 1 & 2 & 4 & 6 & 8 & 10 \\
\hline
  $\Delta_0$ (K) & 1.5 & 3.0 & 3.3 & 10.0 & 14.0 & 20.0 & 23.8 \\
\hline
  $\sigma_0$ (K) & 1.2 & 2.5 & 2.2 & 4.0 & 5.0 & 5.5 & 6.0 \\
\hline
  $\Delta_1$ (K) & 14.0 & 14.5 & 14.7 & 15.0 & 29.0 & 36.0 & 54.0 \\
\hline
  $\sigma_1$ (K) & 4.6 & 5.2 & 2.2 & 11.0 & 9.3 & 18.0 & 19.0 \\
\hline
\hline
\end{tabular}
\end{table}

\begin{table}[htbp]
\caption {The fitting parameters of Eq. (1) and (3), the median of energy gap $\Delta_0$ and $\Delta_1$, the variance $\sigma_0$ and $\sigma_1$ by considering the Gaussian distribution of energy gaps, the nuclear spin energy splitting $\Delta_n$ = 0.1 K, for Tb$_2$Ti$_{1.98}$Zr$_{0.02}$O$_7$ specific heat with increasing magnetic field ($B \parallel$ [111]).} \centering
\renewcommand\arraystretch{1.5}
\begin{tabular}{p{55pt}p{27pt}p{27pt}p{27pt}p{27pt}p{27pt}p{27pt}}
\hline
\hline
  $\mu_0H$ (T) & 0 & 1 & 2 & 4 & 6 & 8  \\
\hline
  $\Delta_0$ (K) & 2.1 & 3.6 & 6.5 & 11.5 & 18.0 & 25.0 \\
\hline
  $\sigma_0$ (K) & 1.9 & 2.5 & 2.8 & 3.5 & 3.7 & 4.5 \\
\hline
  $\Delta_1$ (K) & 15.5 & 15.0 & 20.5 & 26.0 & 32.0 & 43.0 \\
\hline
  $\sigma_1$ (K) & 6.8 & 8.0 & 10.0 & 12.0 & 16.0 & 19.0 \\
\hline
\hline
\end{tabular}
\end{table}

\begin{table}[htbp]
\caption {The fitting parameters of Eq. (1) and (3), the median of energy gap $\Delta_0$ and $\Delta_1$, the variance $\sigma_0$ and $\sigma_1$ by considering the Gaussian distribution of energy gap, the nuclear spin energy splitting $\Delta_n$ = 0.11 K, for Tb$_2$Ti$_{1.9}$Zr$_{0.1}$O$_7$ specific heat with increasing magnetic field ($B \parallel$ [111]).} \centering
\renewcommand\arraystretch{1.5}
\begin{tabular}{p{46pt}p{24pt}p{24pt}p{24pt}p{24pt}p{24pt}p{24pt}p{24pt}}
\hline
\hline
  $\mu_0H$ (T) & 0 & 1 & 2 & 4 & 6 & 8 & 10 \\
\hline
  $\Delta_0$ (K) & 2.8 & 4.8 & 6.0 & 12.0 & 16.5 & 21.0 & 23.0 \\
\hline
  $\sigma_0$ (K) & 2.6 & 3.1 & 4.5 & 6.0 & 9.4 & 14.2 & 14.0 \\
\hline
  $\Delta_1$ (K) & 15.0 & 20.0 & 22.0 & 32.0 & 40.0 & 50.0 & 75.5 \\
\hline
  $\sigma_1$ (K) & 13.0 & 17.0 & 15.0 & 22.0 & 24.0 & 21.0 & 25.4 \\
\hline
\hline
\end{tabular}
\end{table}

\begin{table}[htbp]
\caption {The fitting parameters of Eq. (1) and (3), the median of energy gap $\Delta_0$ and $\Delta_1$, the variance $\sigma_0$ and $\sigma_1$ by considering the Gaussian distribution of energy gap, the nuclear spin energy splitting $\Delta_n$ = 0.11 K, for Tb$_2$Ti$_{1.6}$Zr$_{0.4}$O$_7$ specific heat with increasing magnetic field ($B \parallel$ [111]).} \centering
\renewcommand\arraystretch{1.5}
\begin{tabular}{p{55pt}p{27pt}p{27pt}p{27pt}p{27pt}p{27pt}p{27pt}}
\hline
\hline
  $\mu_0H$ (T) & 0 & 1 & 2 & 4 & 6 & 8 \\
\hline
  $\Delta_0$ (K) & 3.5 & 4.8 & 5.7 & 10.0 & 14.0 & 17.1  \\
\hline
  $\sigma_0$ (K) & 2.4 & 2.9 & 4.7 & 8.9 & 11.5 & 15.9  \\
\hline
  $\Delta_1$ (K) & 41.0 & 45.0 & 47.0 & 55.0 & 68.0 & 75.0  \\
\hline
  $\sigma_1$ (K) & 37.0 & 37.0 & 32.0 & 30.0 & 22.0 & 20.0  \\
\hline
\hline
\end{tabular}
\end{table}

These specific heat data of Tb$_2$Ti$_{2-x}$Zr$_x$O$_7$ ($x =$ 0, 0.02, 0.1 and 0.4) with different magnetic fields are fitted by Eq. (1) and (3) and the fitting parameters are shown in Tables I--IV. The lattice contribution is estimated from that of nonmagnetic Lu$_2$Ti$_2$O$_7$ \cite{JCrystGrowth377}. The energy gaps $\Delta_0$ and $\Delta_1$ are mainly increased with increasing magnetic field, which is consistent with the Zeeman effect. The variance $\sigma_0$ and $\sigma_1$ are also increased with increasing magnetic field, except for the $\sigma_1$ of Tb$_2$Ti$_{1.6}$Zr$_{0.4}$O$_7$. The inhomogeneous distribution of energy splitting widely exists in all samples. The standard deviation is smaller than the mean in all the cases but it is comparable to the mean in some cases, in which the negative part in the integral of Eq. (3) was omitted. The previous inelastic neutron measurements have confirmed the presence of a tetragonal lattice distortion in Tb$_2$Ti$_2$O$_7$ \cite{JPhysConfSer145, Mirebeau1}, which could be the reason of the ground-state doublet splitting. The above analysis of specific heat data indicates that the ground-state doublet has inhomogeneous level splitting not only in Zr-doped samples but also in Tb$_2$Ti$_2$O$_7$. This means that additional contributions other than doping-induced lattice disorder are operative, which may be related to the dynamic Jahn-Teller coupling \cite{Bonville}.

It is worthy of noting that above calculations for specific heat in magnetic fields have taken some simplifications. The first one is about the nuclear Schottky contribution. The nuclear specific heat arises from the non-spinless isotopes present in these samples, including $^{159}$Tb, which is the only isotope present in natural terbium, $^{91}$Zr, which is found in natural zirconium with abundances of 11.22\%, and $^{47}$Ti and $^{49}$Ti, which are found in natural titanium with abundances of 7.4 and 5.4\%, respectively. Since the abundances of non-spinless Zr and Ti isotopes are rather low, we just use $^{159}$Tb to estimate the nuclear specific heat and introduce a parameter $\Delta_n$, which is slightly different for different samples but does not change with magnetic field \cite{Hamaguchi}. The second one is about the CEF contribution. Strictly, the lower-lying four CEF levels must be treated as one system for the specific heat calculation. A magnetic field along the [111] directions lifts the degeneracy of each doublet. The split CEF levels are different between a quarter of Tb$^{3+}$ ions on which the magnetic field is parallel to the local threefold axis and the other Tb$^{3+}$ ions of three quarters. The magnetic-field dependence of the splitting should be determined by the wave functions of two doublets.

\begin{figure}
\includegraphics[clip,width=6.5cm]{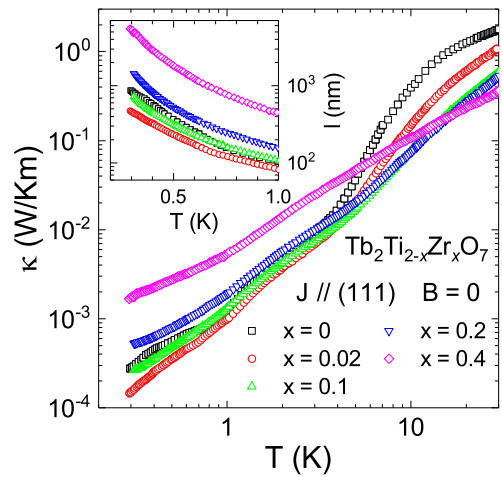}
\caption{Zero-field thermal conductivity of Tb$_2$Ti$_{2-x}$Zr$_x$O$_7$ ($x =$ 0, 0.02, 0.1, 0.2, and 0.4) single crystals. The inset shows the calculated mean free path of phonons.}
\end{figure}

Figure 5 shows the temperature dependence of thermal conductivity ($\kappa$) of Tb$_2$Ti$_{2-x}$Zr$_x$O$_7$ ($x =$ 0, 0.02, 0.1, 0.2, and 0.4) single crystals in zero magnetic field. The magnitudes of $\kappa$ of these crystals are very small, and there is no phonon peak at low temperatures. It is known that the $\kappa(T)$ of insulators usually has an obvious peak at low temperatures (10--20 K), which is a characteristic of phonon heat transport \cite{Berman}. In high-quality single crystals, the absence of phonon peaks is actually very rare. In the previous study of Tb$_2$Ti$_2$O$_7$ crystal, it was found that the mean free path of phonons ($\ell$) is extremely short \cite{Li}. The inset of Fig. 5 shows the $\ell$ at low temperatures for all samples by using the calculation in Ref. \onlinecite{Li}. Even if the temperature is reduced to 0.3 K, $\ell$ is 2--3 orders of magnitude smaller than the sample geometry sizes, which are $\sim$ mm. The microscopic phonon scattering in ordinary single crystals, such as phonon-phonon scattering and the scattering of various crystal defects, will be extinguished at very low temperatures, and the mean free path can reach the size of the sample. This is the so-called the phonon boundary scattering limit \cite{Berman}. It means that at temperature as low as 0.3 K, the phonon scattering in Tb$_2$Ti$_{2-x}$Zr$_x$O$_7$ crystals is still very strong. From the extremely low thermal conductivity of these crystals, it is easy to conclude that if the carriers involved in heat conduction include magnetic excitations, the thermal conductivity of magnetic excitations must be also very small.

The extremely low phonon thermal conductivity of Tb$_2$Ti$_2$O$_7$ has been ascribed to strong scattering of phonon caused by magnetic excitations \cite{Li}. As a candidate for QSL system, Tb$_2$Ti$_2$O$_7$ is likely to have some particular magnetic excitations. First of all, the elementary excitation of QSL, spinon, can either transport heat or scatter phonon or both. The recent thermal Hall effect results indicated the possible spinon excitations in this material \cite{Ong}. Alternatively, the neutron scattering has discovered that in the spin liquid state of Tb$_2$Ti$_2$O$_7$ the excited CEF level is strongly coupled to the transverse acoustic phonons, forming a hybrid excitation \cite{Fennell}. Since the energy gap between the ground-state doublet and the first excited level is not a constant, suggested by the above specific data, the hybrid excitation may occur in a broad range of temperature. In addition, the splitting of the ground-state doublet is also not a constant, which will result in further coupling or scattering between acoustic phonons and CEF levels. This can be another understanding of the strong phonon scattering by magnetic excitations. Ruminy {\it et al.} reported that the phonon band structure of pyrochlores have essentially identical features, with adjustments that can be classified in two ways \cite{Ruminy}. First, the larger ionic radius of both the $A$- and $B$-site ions lead to an expansion of the unit cell, which reduces the frequencies of phonon vibrations across the entire phonon spectrum. Second, the larger the mass of the $B$-site ion, the more the statistical weight of its partial phonon density of states will be shifted to lower frequencies. In Zr-doped samples, the frequencies of phonon vibrations shift to lower frequency because the Zr$^{4+}$ ion has larger ionic radius and larger mass than Ti$^{4+}$ ion. In lightly doped samples ($x =$ 0.02 and 0.1), the coupling between phonons and CEF levels is likely stronger than Tb$_2$Ti$_2$O$_7$ and the thermal conductivity decreases. With further doping Zr ($x =$ 0.2 and 0.4), the phonon energy decreases and the spin-phonon coupling is weakened, which results in the recovery of thermal conductivity at low temperatures. Except for the magnetic excitations, the scattering of phonon caused by lattice disorder should be considered in Zr-doped samples. It is notable that the high-$T$ $\kappa$ ($T >$ 10 K) is reduced with Zr doping, which is mainly due to the stronger lattice disorder scattering on phonons.

\begin{figure}
\includegraphics[clip,width=8.5cm]{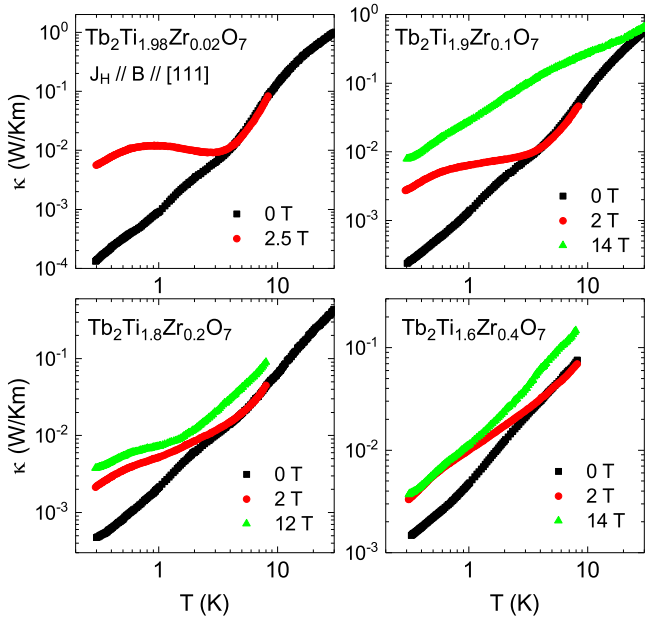}
\caption{Temperature dependence of thermal conductivity of Tb$_2$Ti$_{2-x}$Zr$_x$O$_7$ ($x =$ 0.02, 0.1, 0.2, and 0.4) single crystals under different magnetic fields. Both magnetic field and heat current are applied along the [111] direction.}
\end{figure}

Figure 6 shows temperature dependence of thermal conductivity of Tb$_2$Ti$_{2-x}$Zr$_x$O$_7$ ($x =$ 0.02, 0.1, 0.2, and 0.4) single crystals for magnetic field and heat current along the [111] direction. At low temperatures, the $\kappa$ is strongly enhanced in magnetic field, which is likely due to magnetic field suppressing of the magnetic excitations and, therefore, weakening the phonon scattering. The following results of the magnetic field dependence of $\kappa$ display more details.

\begin{figure}
\includegraphics[clip,width=8.5cm]{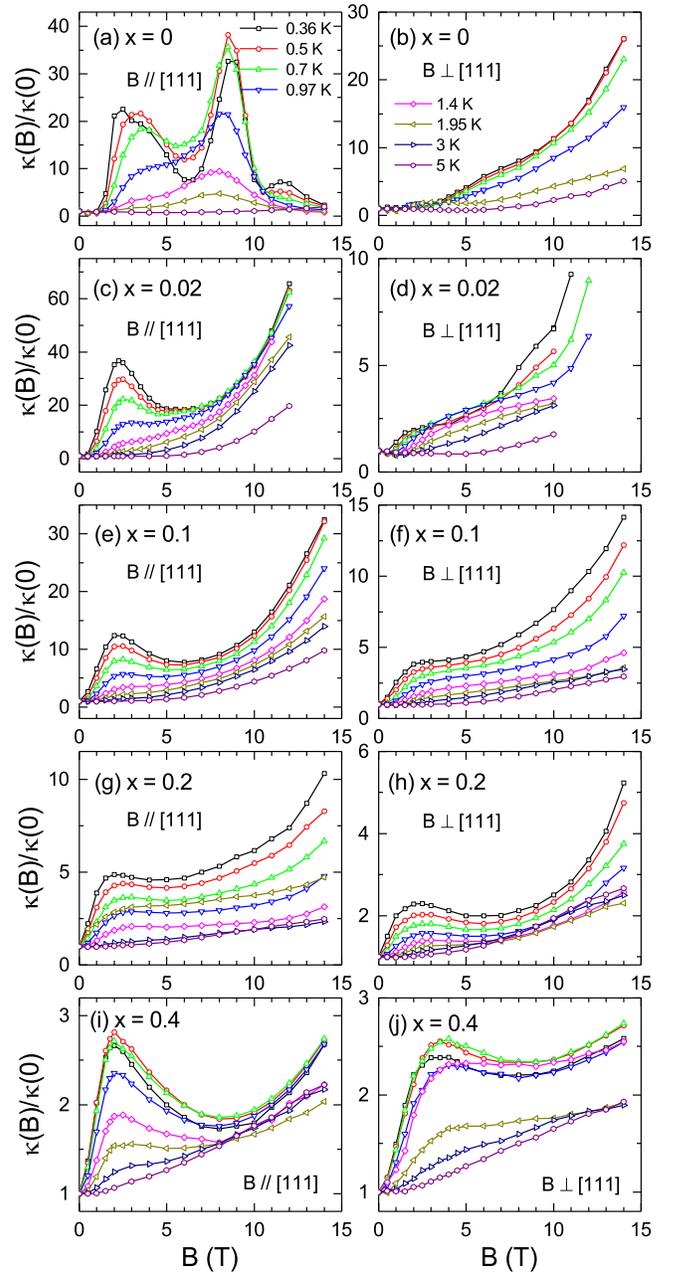}
\caption{Magnetic field dependence of thermal conductivity of Tb$_2$Ti$_{2-x}$Zr$_x$O$_7$ ($x =$ 0, 0.02, 0.1, 0.2, and 0.4) single crystals. The heat current is along the [111] direction while the magnetic field is either parallel to or perpendicular to it.}
\end{figure}

Figure 7 shows the magnetic-field dependence of $\kappa$ at low temperatures for Tb$_2$Ti$_{2-x}$Zr$_x$O$_7$ ($x =$ 0, 0.02, 0.1, 0.2, and 0.4) single crystals with magnetic fields along or perpendicular to the [111] direction. For $B \parallel$ [111], the 0.36 K $\kappa(B)$ curve of the $x =$ 0 sample shows three peaks (at 2.5, 8.5 and 11.5 T) and three dips (at 0.5, 6 and 10.5 T), which has been reported in our earlier work \cite{Li}. In principle, the dip in $\kappa(B)$ is likely related to some field-induced magnetic transition \cite{Sun_DTN, Zhao_IPA, Wang_HMO, Wang_TMO, Zhao_GFO, Zhao_DFO, PhysRevB1062022}. However, it is hard to image so complicated magnetic transitions in Tb$_2$Ti$_2$O$_7$ if each dip on the $\kappa(B)$ curve corresponds to a magnetic transition. A neutron-scattering experiment had found that the magnetic field along the [111] direction could induce an AF order \cite{Y.Yasui}. The elastic neutron-scattering intensity was found to increase with field up to 2--3 T and became nearly saturated, which may have some correspondence with the sharp increase of $\kappa$ at low field. However, the high-field anomalies in the $\kappa(B)$ cannot be related to some magnetic transitions. This complicated $\kappa(B)$ behavior also cannot be simply explained with spinon scenario, in which the spinons are usually suppressed in high magnetic fields. One possible origin is the field-induced change of phonon scattering by CEF levels, which are further split by Zeeman effect. Zr doping induces some significant change of the $\kappa(B)$ behavior. First, it is obvious that two peaks at high field disappear in Zr-doped samples. Second, the $\kappa$ keeps increasing at high magnetic fields. Third, the low-field peak weakens with increasing Zr content, which may be due to Zr suppressing the magnetic excitations. In addition, with increasing temperature the low-field peak gradually disappears.

For $B \perp$ [111], the very-low-$T$ $\kappa$ of the $x = $0 sample is nearly field independent at $B <$ 2 T and then increases monotonically with field. The effect of magnetic field along the [110] direction on magnetism has been well studied \cite{PhysRevLett96, PhysRevLett101} and the Tb$^{3+}$ spins were thought to be lying on two sets of chains along the [110] and [1\=10] directions. The spins on chains parallel to the field direction (the so-called $\alpha$ chains) align along the local [111] axis with a component parallel to the field direction. Whereas the spins on chains perpendicular to the field direction (the $\beta$ chains) favor an AF order at high fields. Elastic neutron scattering intensity was found to increase gradually above 2 T and at subkelvin temperatures \cite{PhysRevLett96, PhysRevLett101}, which indicates the field-induced order for field along the [110] direction and has some correspondence to the increase of $\kappa(B)$ above 2 T. In the $x =$ 0.02 and 0.1 samples, a shoulder-like feature of $\kappa(B)$ shows up at about 2 T and shifts to higher magnetic field with increasing temperature. This may be related to that Zr doping changes the crystal environment and has an effect on spin orientation of Tb$^{3+}$ on the $\alpha$ and $\beta$ chains. In higher-doping samples, the shoulder-like feature evolves into a broad peak at about 3 T.

\begin{figure}
\includegraphics[clip,width=8.5cm]{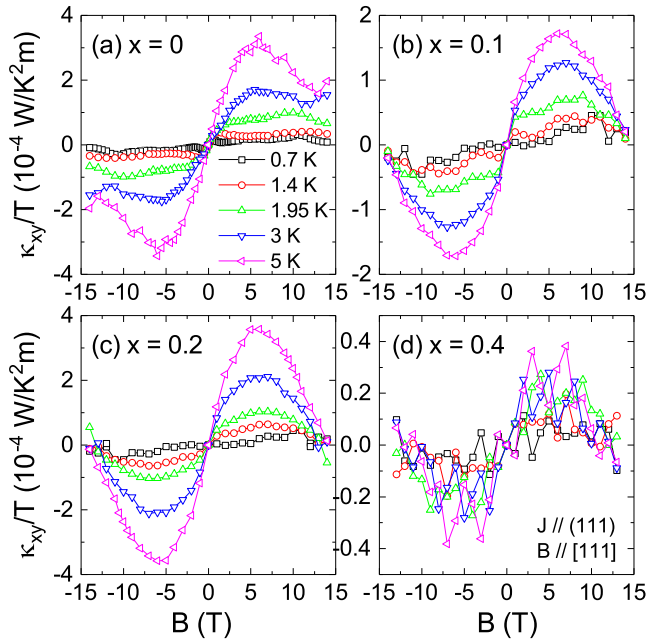}
\caption{Magnetic field dependencies of the thermal Hall conductivity of Tb$_2$Ti$_{2-x}$Zr$_x$O$_7$ ($x =$ 0, 0.1, 0.2, and 0.4) single crystals for $B \parallel$ [111].}
\end{figure}

These thermal conductivity results are still lacking of quantitative explanation. It seems that they may be closely related to the change of CEF levels with Zr doping and increasing field, that affect the coupling between acoustic phonons and CEF levels. However, from these data it is not clear whether the possible spinon excitations in the QSL state take part in the heat transport, either carrying heat or scattering phonons. The following thermal Hall results can provide the information on the role of spinons.

Figure 8 shows the thermal Hall conductivity of Tb$_2$Ti$_{2-x}$Zr$_x$O$_7$ ($x =$ 0, 0.1, 0.2, and 0.4) single crystals at different temperatures with magnetic field parallel to the [111] direction. Note that the present experimental results confirm the large thermal Hall effect in this material. Actually, our $\kappa_{xy}(B)$ data are several times larger than the previous work, which also indicates the high quality of our single crystals. The $x =$ 0 sample exhibits magnetic field dependence of $\kappa_{xy}$ similar to that reported by Hirschberger {\it et al.} \cite{Ong} In general, the $\kappa_{xy}(B)$ increases with increasing field at low fields and shows a broad peak around 8 T. Since our measurements were carried out under much higher fields (up to 14 T) than the previous work, one can find that the $\kappa_{xy}(B)$ significantly decreases at high fields. This non-monotonic behavior of $\kappa_{xy}(B)$ is important for understanding the origin of such a large thermal Hall effect. Two possible origins can be discussed. First, the non-monotonic $\kappa_{xy}(B)$ behavior is compatible with the spinon origin proposed by Hirschberger {\it et al.} \cite{Ong} It should be pointed out that Tb$_2$Ti$_2$O$_7$ has extremely small thermal conductivity in zero field. This means that both the phonons and spinons have very weak ability of transporting heat at low temperatures, which was ascribed to the strong scattering between phonons and magnetic excitations. Thus, one may ask whether it is reasonable that the spinons can exhibit large thermal Hall conductivity in this material. Figure 6 shows that the thermal conductivity is strongly enhanced by applying magnetic field. Apparently, the spin-phonon scattering is significantly suppressed by magnetic field and therefore both the phonons and spinons have larger ability of transporting heat with increasing field. Thus, the large $\kappa_{xy}$ which increases quickly with field can be due to the transport of spinons. However, at high magnetic fields the spinon excitations would be significantly suppressed, which results in the decrease of $\kappa_{xy}$. Second, the $\kappa_{xy}(B)$ behavior may be due to the resonant skew scattering of phonons from the crystal field levels of Tb$^{3+}$ ions, which was proposed from the phonon thermal Hall effect in  Tb$_3$Gd$_5$O$_{12}$ \cite{Mori, Wawrzynczak}. This scenario would be also rather promising for Tb$_2$Ti$_2$O$_7$ since the CEF levels significantly contribute to the thermodynamics, as the specific heat data indicate. In this case, the broad peak at 8 T is related to the large energy gap between the ground CEF level and others. Further investigations are called for clarifying this interesting phenomenon.

In the Zr-doped samples with $x =$ 0.1 and 0.2, the thermal Hall results are essentially the same as that of undoped sample. Since the field dependence of thermal conductivity is quite different between these Zr-doped samples and undoped sample, it is likely that the thermal conductivity behavior is dominated by the phonons while the thermal Hall conductivity cab be determined by spinons. In highly doped sample with $x =$ 0.4, the $\kappa_{xy}(B)$ still show similar behavior with other samples but its value is one order of magnitude smaller, which may be related to the strong scattering effect by the disorder.

\section{SUMMARY}

In summary, we grew the single crystals of Tb$_2$Ti$_{2-x}$Zr$_x$O$_7$ ($x =$ 0, 0.02, 0.1, 0.2, and 0.4) and studied their specific heat, thermal conductivity and thermal Hall effect at low temperatures and in high magnetic fields. The magnetic specific heat are quantitatively analyzed by using the modified Schottky formula, considering a Gaussian distribution of the energy split of the ground-state doublet and the gap between the ground state and first excited level. At low temperatures, these crystals show the extremely low thermal conductivity with strong magnetic field dependence, indicating strong scattering between phonons and magnetic excitations. The thermal Hall conductivity $\kappa_{xy}(B)$ are large at low temperatures and display a broad peak at 8 T. The disappearance of $\kappa_{xy}(B)$ signal at higher fields demonstrate its origin of either the spinon transport or the phonon skew scattering.

\begin{acknowledgements}

We thank J. D. Song, X. Rao, and W. J. Chu for helps on experiments. This work was supported by the National Natural Science Foundation of China (Grant Nos. 11874336, U1832209, 12274388, 12174361, 12104010, and 12104011) and the Nature Science Foundation of Anhui Province (Grant Nos. 1908085MA09 and 2108085QA22).

\end{acknowledgements}

\end{document}